\documentclass[twocolumn,aps,showpacs,amsmath,amssymb]{revtex4}

\usepackage{graphicx}
\usepackage{dcolumn}
\usepackage{bm}

\begin{document}

\title{
Nucleus-nucleus potential with shell-correction contribution and deep sub-barrier fusion of heavy nuclei
}%

\author{
V. Yu. Denisov
}%

\affiliation{%
Institute for Nuclear Research, Prospect Nauki 47,
03680 Kiev, Ukraine \\
Faculty of Physics, Taras Shevchenko National University of Kiev, Prospect Glushkova 2, 03022 Kiev, Ukraine
}%

\date{\today}

\begin{abstract}
It is suggested that the full nucleus-nucleus potential consists of the macroscopic and shell-correction parts. The deep sub-barrier fusion hindrance takes place in a nucleus-nucleus system with a strong negative shell-correction contribution to the full heavy-ion potential, while a strong positive shell-correction contribution to the full potential leads to weak enhancement of the deep sub-barrier fusion cross section.
\end{abstract}


\maketitle

\section{Introduction}

The nucleus-nucleus potential is a key ingredient in the description of nuclear reactions \cite{bass80,satchler,fl,dp}. Microscopic evaluation of the potential between nuclei is based on both the effective nucleon-nucleon interaction and the nucleon density distributions of interacting nuclei, see Refs. \cite{satchler,fl,dp,bbk,bs,m3y,dn,dnest1,uo,uomr} and papers cited therein. There are also simple parametrizations of the nucleus-nucleus potential \cite{bass80,fl,dp,bass73,prox77,kns,aw,winther,prox2000,d2002,sww,duttp}, which are often used to describe various heavy-ion reactions.

Various nucleus-nucleus potentials have been applied to the description of heavy-ion fusion around barrier. The one-dimensional barrier penetration model with potentials, which fit to the fusion cross sections at energies exceeding the barrier height, leads to strong underestimation of sub-barrier fusion cross sections \cite{fl,dp,beckerman,dhrs,bt,ht}. It is found that the couplings to both the low-energy surface vibrational states and the nucleon transfer between nuclei greatly enhance the sub-barrier fusion cross sections \cite{fl,dp,beckerman,dhrs,bt,ht,dtr}. The heavy-ion fusion cross sections around the barrier are well described in the framework of such coupled-channel models.

Further detailed experimental studies show that there is fusion hindrance at deep sub-barrier energies \cite{sbfh1,sbfh2,OPb,CaCa,NiFe}. In this case the values of measured fusion cross sections at deep sub-barrier energies are smaller than the ones evaluated in the coupled-channel models, which well describe the fusion data at around barrier energies. Researchers have tried to explain this phenomenon is tried to explain in the framework of various approaches, see, for details, Refs. \cite{rep,ihi,saasz,im,sdhw} and papers cited therein.

The goal of the present study is to discuss a new mechanism of fusion hindrance at deep sub-barrier energies that is related to the microscopic shell-correction contribution to the full nucleus-nucleus potential. According to the Strutinsky shell-correction prescription \cite{s} the microscopic shell-correction energies should be added to the macroscopic contribution for the correct evaluation of the full energy of a nuclear system. Therefore the full nucleus-nucleus potential consists of the macroscopic and shell-correction parts.

When nuclei are approaching each other the shell structures of both nuclei are changed due to the interaction of nucleons belonging to different nuclei. The energies of nucleon single-particle levels of each nucleus are shifted and the single-particle levels are split due to the interaction between nucleons in different nuclei. Therefore the microscopic shell-correction energies \cite{s} of both nuclei are changed at small distances between nuclei. The variations of shell-correction energies induced by the interaction of nucleons belonging to different nuclei lead to modification of the full interaction potential between nuclei. I propose a simple approach for evaluation of the shell-correction contribution to the full nucleus-nucleus potential.

I describe the fusion cross section data for reactions $^{16}$O+$^{208}$Pb \cite{OPb}, $^{48}$Ca+$^{48}$Ca \cite{CaCa} and $^{58}$Ni+$^{54}$Fe \cite{NiFe} using the full nucleus-nucleus potential consisting of the macroscopic and microscopic parts. I select these reactions because the experimental fusion cross sections for them are known in very wide energy ranges. The couplings to the low-energy surface vibrational states are taken into account when evaluating of the fusion cross sections. I analyze the influence of the microscopic shell-correction contribution to the full potential on the heavy-ion fusion cross sections and show that this contribution has a strong effect on the cross sections deeply below the barrier. A detailed description of this approach for the nucleus-nucleus potential is presented in Sec. 2. Sec. 3 is a discussion of the results and gives the conclusions.

\section{Nucleus-nucleus potential with shell-correction contribution}

Applying the Strutinsky shell-correction prescription \cite{s} to the system of interacting nuclei, one gets the full interaction potential in the form (see also \cite{dr96,iims,dclust})
\begin{eqnarray}
V_{\rm tot}(R) &=& V_{\rm macro}(R) + V_{\rm sh}(R) \nonumber \\ &=& [E_{12}(R) - E_1 - E_2 ] \nonumber \\ &+& [\delta E_{12}(R) - \delta E_1 - \delta E_2].
\end{eqnarray}
Here $E_1$, $E_2$, $ \delta E_1$ and $\delta E_2$  are the macroscopic and shell-correction energies of the non-interacting spherical nuclei 1 and 2, respectively, $E_{12}(R)$ and $\delta E_{12}(R)$ are the macroscopic and shell-correction energies of interacting nuclei at distance $R$ between the mass centers of separated nuclei, correspondingly. Shell-correction energies $\delta E_{12}(R)$, $ \delta E_1$ and $\delta E_2$ include the proton and neutron shell-correction energies related to both the non-uniformity of single-particle spectra around the Fermi energies and the pairing corrections \cite{s}.

According to the Strutinsky shell-correction prescription the macroscopic energy of a nuclear system is parametrized by a simple expression of the liquid-drop model. Similarly, I propose that the macroscopic part of the nucleus-nucleus potential at large distances can be described by the sum of the Woods-Saxon nuclear potential and the Coulomb potential. The Woods-Saxon potential is introduced to describe the interaction between the nucleon and the nucleus with diffused distribution on nucleon density. The interaction of two nuclei in the framework of a single-folding potential is related to the nucleon-nucleus interaction, therefore presentation of  the macroscopic interaction in the sum of the Woods-Saxon and the Coulomb parts is reasonable. Moreover, the Woods-Saxon potential is very widely and very successfully used to describe of various nucleus-nucleus reactions. It is also often used for the description of sub-barrier heavy-ion fusion \cite{beckerman,dhrs,bt,ht,OPb,CaCa,NiFe,rep,ihi,ccfull}. Detailed description of the macroscopic Woods-Saxon part of the full potential will be given later.

\subsection{Microscopic part of the full potential}

It is obvious that the mutual influence of nuclei on their single-particle spectra is negligible at distances $R$ much greater than $R_t=R_1+R_2$, where $R_i=r_0 A^{1/3}_i$ is the radius of a nucleus with $A_i$ nucleons ($i=1,2$). Therefore, we can write in this limit
\begin{eqnarray}
\delta E_{12}(R)|_{R \gg R_t} = \delta E_1 + \delta E_2 .
\end{eqnarray}

Value of $\delta E_{12}(R)$ at $R = 0$ equals the shell-correction energy of the ground-state nucleus formed at complete fusion of nuclei 1 and 2. The full potential should be equal to the fusion reaction $Q$-value at $R=0$, i.e. $V_{\rm tot}(0)=Q$.

The nuclei strongly interact at small distances between them. This interaction leads to the shift and splitting of the single-particle levels in both nuclei. Due to this the proton and neutron single-particle spectra around the Fermi levels become more homogenous around distance $R_{t}$. Such behavior of single-particle levels of two nuclei is clearly demonstrated in the framework of the two-center shell model \cite{msd,mg} and the time-dependent Hartree-Fock calculations \cite{uomr}.

The absolute value of shell-correction energy is reduced in the case of more homogenous single-particle spectra around the Fermi levels \cite{s}. A sharp reduction of the shell-correction contribution to the full potential energy around the touching point of the nuclei is obtained in Ref. \cite{msd}. So, the energy level splitting and shifts, which are proportional to the strength of mutual nucleus-nucleus perturbation, reduce the values of shell-correction energies of nuclei at small distances between them.

The perturbation of single-particle levels is enlarged when the distance decreases between the surfaces of the nuclei and when interaction increases between nucleons belonging to different nuclei. The perturbation strength is related to the density distribution in the nucleus induced by the disturbance as well as the radius of nucleon-nucleon force. The density distribution is often parametrized by the Fermi distribution, therefore we approximate the shell-correction contribution to the full nucleus-nucleus potential at small and large distances between nuclei as \cite{dclust}
\begin{eqnarray}
V_{\rm sh}^0(R) &=& \delta E_{12}(R) - \delta E_1 - \delta E_2 \nonumber \\ &=& [\delta E_1 + \delta E_2 ] f_{\rm sh}(R)- \delta E_1 - \delta E_2 \nonumber\\ &=& [\delta E_1 + \delta E_2 ] (f_{\rm sh}(R)-1).
\end{eqnarray}
where
\begin{eqnarray}
f_{\rm sh}(R) = 1/\left\{1 + \exp{[(R_{\rm sh}-R)/d_{\rm sh}}]\right\},
\end{eqnarray}
and $R_{\rm sh}=r_{0}^{\rm sh} (A_1^{1/3}+A_2^{1/3})$ and $d_{\rm sh}$ are the radius and diffuseness related to the attenuation of the shell-correction with reduction of distance $R$. My approximation for shell-correction energy contribution into the full potential is rough, but it can greatly simplify the calculations of shell-correction energies for various nucleus-nucleus systems at different distances $R$. Note that exponential reduction of shell-correction energy values related to washing out the shell non-homogeneity of single-particle spectra is often considered in nuclear physics \cite{dp,ms,ist,dh}.

The shell-correction energy of interacting nuclei $\delta E_{12}(R)$ is smoothly approached to the limit of non-interacting nuclei (2) at large distances $R$ between nuclei, i.e. $V_{\rm sh}^0(R) \rightarrow 0$ at $R \gg R_{\rm sh} \sim R_t$.

It is possible to evaluate the shell-correction energy values $\delta E_1$ and $\delta E_2$ according to the Strutinsky prescription \cite{s} for a specific nucleon mean-field approximation. However, the easiest way to estimate the value of shell-correction energy $\delta E$ in a spherical nucleus is to find the difference between the experimental $B_{\rm exp}$ and macroscopic $B_{\rm m}$ binding energies of the nucleus
\begin{eqnarray}
\delta E = B_{\rm exp} - B_{\rm m}.
\end{eqnarray}
This expression corresponds to the physical sense of the shell-correction energy.
The values of $B_{\rm exp}$ can be found in the recent evaluation of the atomic masses \cite{ame2012}, while the value of $B_{\rm m}$ in nucleus with $Z$ protons and $N$ neutrons is
\begin{eqnarray}
B_{\rm m} = -15.86864 A+21.18164 A^{2/3}-6.49923 A^{1/3} + \nonumber \\
 \left[\frac{N-Z}{A}\right]^2 \left[26.37269 A -23.80118 A^{2/3} - 8.62322 A^{1/3} \right] \nonumber \\
 + \frac{Z^2}{A^{1/3}} \left[ 0.78068- 0.63678 A^{-1/3} \right] \nonumber \\ + P_p+P_n.
 \;\;\;
\end{eqnarray}
Here $B_{\rm m}$ is the binding energy in MeV, $A=Z+N$ are the number nucleons in the nucleus, $P_{p(n)}$ are the proton (neutron) pairing terms, which equal to $P_{p(n)}=5.62922 (4.99342) A^{-1/3}$ in the case of odd $Z$ ($N$) and $P_{p(n)}=0$ in the case of even $Z$ ($N$). Equation (6) is a simple extension of the Weizs\"acker formula for the binding energy of nuclei. I obtained the values of the coefficients in Eq. (6) by fitting the recent values of the atomic masses \cite{ame2012}. The experimental binding energies of 3353 nuclei are described by Eq. (6) with a root mean error of 2.49 MeV. This error is very small compared to the experimental values of atomic binding energies in medium and heavy nuclei. Note that the  shell-correction energy evaluation technique used here is similar to the one applied to evaluate the energy level density in nuclei; see, for example, \cite{ripl3}.

I have ignored any influence of the relative motion of nuclei on the nucleus-nucleus interaction and single-particle levels up to now. However, the nucleons move in the approaching nuclei during reaction and interaction of nucleons belonging to different nuclei disturbs the nucleons in both nuclei. This leads to the disturbance of the shell structure of each nucleus. It is obvious that the strength of this disturbance should depend on the ratio between the nucleon velocity (energy) and the relative velocity (kinetic energy) of the nuclei. When the approaching nuclei slow down, the static (adiabatic) consideration is close to realistic, because the shell structure of nuclei can adjust to the static one for the corresponding distance between nuclei. In contrast to this, the shell structure and nucleon density distributions cannot be disturbed at fast collisions; therefore nuclei can touch each other without modification of the shell structure of colliding nuclei at high collision energies. Therefore, the nuclei overcome the barrier essentially in their ground-state density at high collision energies, see also \cite{wl,umar}. In order to take into account this effect I introduce the dependence of the shell-correction energy contribution to the full potential on the collision energy $E$
\begin{eqnarray}
V_{\rm sh}(R,E) = \left\{ \begin{array}{lr} V_{\rm sh}^0(R) \exp{[-a(E-B)]}, & {\rm at} \; E \geq B, \\
V_{\rm sh}^0(R), & {\rm at} \; E \leq B, \\
\end{array} \right.
\end{eqnarray}
where $B$ is the barrier height of the macroscopic potential $V_{\rm macro}(R)$ and $a$ is the reduction parameter of the shell-correction energy contribution to the full potential. So, the shell-correction energy contribution to the full potential decreases with increasing collision energy. Equation (7) does not depend on the velocity of the nucleons, because nucleon velocity at the Fermi level has a small variation from one nucleus to another and the value of the shell-correction energy depends on the inhomogeneity of the single-particle spectra around the Fermi energy \cite{s}.

\subsection{
Full nucleus-nucleus potential
}

The full nucleus-nucleus potential is the sum of Coulomb and nuclear parts
\begin{eqnarray}
V_{\rm tot}(R,E) = V_{\rm C}(R) + V_{\rm nucl}(R,E)),
\end{eqnarray}
where
\begin{eqnarray}
V_{\rm C}(R) = \left\{ \begin{array}{lr}
\frac{Z_1 Z_2 e^2}{R}, & R \geq R_{\rm C}, \\
\frac{Z_1 Z_2 e^2}{R_{\rm C}} \left[ \frac{3}{2} -\frac{1}{2} \; \frac{R^2}{R_{\rm C}^2} \right], & 0 \leq R \leq R_{\rm C} , \\
\end{array} \right.
\end{eqnarray}
\begin{eqnarray}
V_{\rm nucl}(R,E) = \left\{ \begin{array}{lr}
V_{\rm macro}^{\rm out}(R) + V_{\rm sh}(R,E), & R \geq R_{\rm m}, \\
V_{\rm nucl}^{\rm in}(R), & 0 \leq R \leq R_{\rm m}. \\
\end{array} \right.
\end{eqnarray}
Equation (9) for Coulomb potential is standard for heavy-ion reactions, see for example \cite{bass80,satchler,fl,dp}. Let's put $R_{\rm C}=R_{\rm t}$ for the sake of the reduction of the parameter number.

The macroscopic nuclear part $V_{\rm macro}^{\rm out}(R)$ of the potential at large distances is taken in the Woods-Saxon form
\begin{eqnarray}
V_{\rm macro}^{\rm out}(R) = - \frac{R_1 R_2}{R_t} \; \frac{V_0}{1+\exp[(R-R_{\rm t})/d]} ,
\end{eqnarray}
where $V_0$ and $d$ are the parameters of strength and diffuseness, respectively.

The nuclear part of the potential at small distances is taken in the Woods-Saxon form too
\begin{eqnarray}
V_{\rm nucl}^{\rm in}(R)=\frac{Q_{\rm eff}}{1+\exp[(R-R_{\rm in})/d_{\rm in}]} ,
\end{eqnarray}
where
\begin{eqnarray}
Q_{\rm eff} = Q - \frac{3Z_1 Z_2 e^2}{2R_{\rm C}}, \\
d_{\rm in} =-\frac{V_{\rm nucl}^2(R_{\rm m})}{Q_{\rm eff} V_{\rm nucl}^\prime(R_{\rm m})} \left(\frac{Q_{\rm eff}}{V_{\rm nucl}(R_{\rm m})}-1 \right) , \\
R_{\rm in}= R_{\rm m}-d_{\rm in} \log\left(\frac{Q_{\rm eff}}{V_{\rm nucl}(R_{\rm m})}-1 \right) .
\end{eqnarray}
Here $Q$ is the $Q$-value of the fusion reaction evaluated by using the recent values of the atomic masses presented in Ref. \cite{ame2012}. If the binding energy of the nucleus is not given in Ref. \cite{ame2012}, then it can be obtained with the help of Eq. (6). The radius $R_{\rm in}$ and diffuseness $d_{\rm in}$ of the inner nuclear potential are obtained by using the continuity conditions of the potential and its derivative at the matching point $R_{\rm m}=R_{\rm t}$, i.e.
\begin{eqnarray}
V_{\rm nucl}(R_{\rm m})&=&V_{\rm macro}^{\rm out}(R_{\rm m}) + V_{\rm sh}(R_{\rm m},E) \nonumber \\&=& V_{\rm nucl}^{\rm in}(R_{\rm m}), \\
V_{\rm nucl}^\prime (R_{\rm m})&=&\frac{d}{dR}[V_{\rm macro}^{\rm out}(R_{\rm m}) + V_{\rm sh}(R_{\rm m},E)] \nonumber \\&=& \frac{d}{dR}[V_{\rm nucl}^{\rm in}(R_{\rm m})].
\end{eqnarray}
In general, the position of the matching point may not coincide with $R_{\rm t}$. Such a position of the matching point is chosen here for the sake of reduction of the parameter number.

The nucleus-nucleus interaction at large distances $R \geq R_{\rm t}$ contains the macroscopic nuclear, Coulomb and shell-correction energy terms, see Eqs. (8)--(11). The macroscopic nuclear part of the interaction has the shape of a Woods-Saxon potential, see Eq. (10). The shell-correction contribution is related to the attenuation of shell-correction energies in both nuclei due to interaction between them and can be found with the help of Eqs. (3)--(7) and recent data for the atomic masses \cite{ame2012}.

According to Eq. (1) the full potential $V_{\rm tot}(0) = V_{\rm nucl}^{\rm in}(0)+\frac{3Z_1 Z_2 e^2}{2R_{\rm t}}$ is approximately equal to $Q$ due to Eqs. (8)--(10), (12)--(15). I have chosen the Woods-Saxon form of the nuclear part of potential, see Eqs. (11) and (12), because the Woods-Saxon potential is very common in nuclear reaction theory. Therefore the proposed potential can be easily integrated into various existing codes for description of different properties of nuclear reactions.

\section{Fusion cross sections}

I evaluate the fusion cross sections of nuclei  by using code CCFULL \cite{ccfull}. The coupling to the low-energy surface $2^+$, $3^-$, $2^+ \otimes 3^-$ vibrational states as well as all possible mutual couplings are taken into account in both nuclei at my coupled-channel calculation of fusion cross sections. The energies and deformation parameters of $2^+$ and $3^-$ states, which are necessary for evaluation of the fusion cross sections, are taken from Refs. \cite{be2,be3}.

My coupled-channel calculations are simple and may be extended by taking into account additional channels and couplings. The goal of this study is to show the influence of the shell-correction energy contribution into the full heavy-ion potential on the fusion cross sections. Therefore I take to account only pointed coupled channels and couplings, which give the most important contribution into sub-barrier fusion cross sections for reactions considered below.

\begin{figure}
\begin{center}
\includegraphics[width=8.7cm]{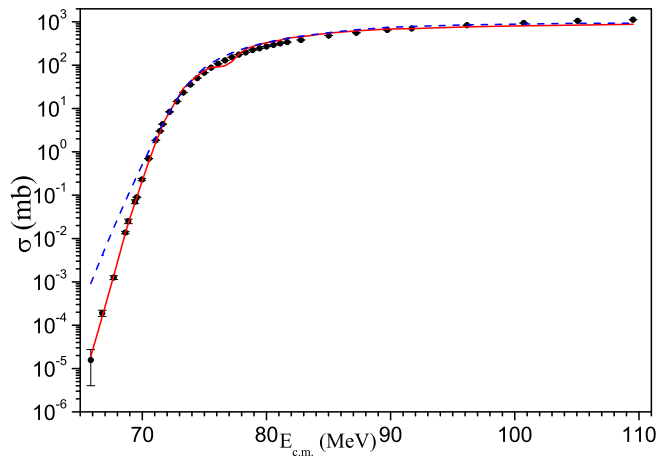}
\caption{(Color online) Fusion cross sections evaluated by using both the macroscopic and shell-correction parts of the nucleus-nucleus potential (solid line) compared with the experimental data (dots) for the reaction $^{16}$O+$^{208}$Pb \cite{OPb}. The fusion cross sections obtained with the help of the macroscopic part on the nucleus-nucleus potential is shown by the dashed line.}
\end{center}
\end{figure}

The fusion cross sections evaluated by using the full (the sum of macroscopic and shell-correction energy parts) nucleus-nucleus potential are compared with the experimental data for reaction $^{16}$O+$^{208}$Pb \cite{OPb} in Fig. 1. The values of potential parameters are presented in Table 1. The results of calculation with the full potential agrees well with the data. In contrast to this, the results obtained with the macroscopic part of the nucleus-nucleus potential overestimate the cross section at very low collision energies, see Fig. 1. So, the shell-correction contribution to the nucleus-nucleus interaction is very important and can explain the fusion hindrance at deep sub-barrier energies.

\begin{table}
\caption{The values of potential parameters.}
\begin{tabular}{ccccccc}
\hline \hline
Reaction & $V_0$ (MeV/fm) & $r_0$ (fm) & $d$ (fm) & $r_{0}^{\rm sh}$ (fm) & $d_{\rm sh}$ (fm) \\
\hline
$^{16}$O+$^{208}$Pb & 18.462 & 1.2270 & 0.54592 & 1.1826 & 0.21970 \\
$^{48}$Ca+$^{48}$Ca & 15.719 & 1.2307 & 0.60868 & 1.2041 & 0.27687 \\
$^{58}$Ni+$^{54}$Fe & 19.607 & 1.2288 & 0.58794 & 1.2016 & 0.20269 \\
\hline \hline
\end{tabular}
\end{table}

The values of shell-correction energies of $^{16}$O and $^{208}$Pb are $-5.249$ and $-10.620$ MeV, respectively. The fusion cross sections for this reaction are measured in a very wide energy range \cite{OPb}. Taking into account the large amplitudes of shell-correction energies in $^{16}$O and $^{208}$Pb and the wide collision energy range for this reaction, one can fix the value of parameter $a=0.35$ MeV$^{-1}$ related to the reduction of the shell correction contribution into the total heavy-ion potential, see Eqs. (7), (8), and (10).

The full (the sum of Coulomb, nuclear, and shell-correction parts) and macroscopic (the sum of Coulomb and nuclear parts) potentials for the $^{16}$O+$^{208}$Pb system at $E=B$ are presented in Fig. 2. The shell-correction energy contribution slightly changes the potential at distances a little smaller than the barrier radius. However this contribution is noticeable at distances around the touching point ($R_t=10.36$ fm) and the capture well. The full potential is more shallow than the macroscopic potential.

The contribution of the shell-correction energy to the full potential depends on collision energy $E$, therefore the couplings related to low-energy vibrational states vary with $E$. The full potential at $E>B$ smoothly tends towards the macroscopic potential with increasing $E$.

\begin{figure}
\begin{center}
\includegraphics[width=8.7cm]{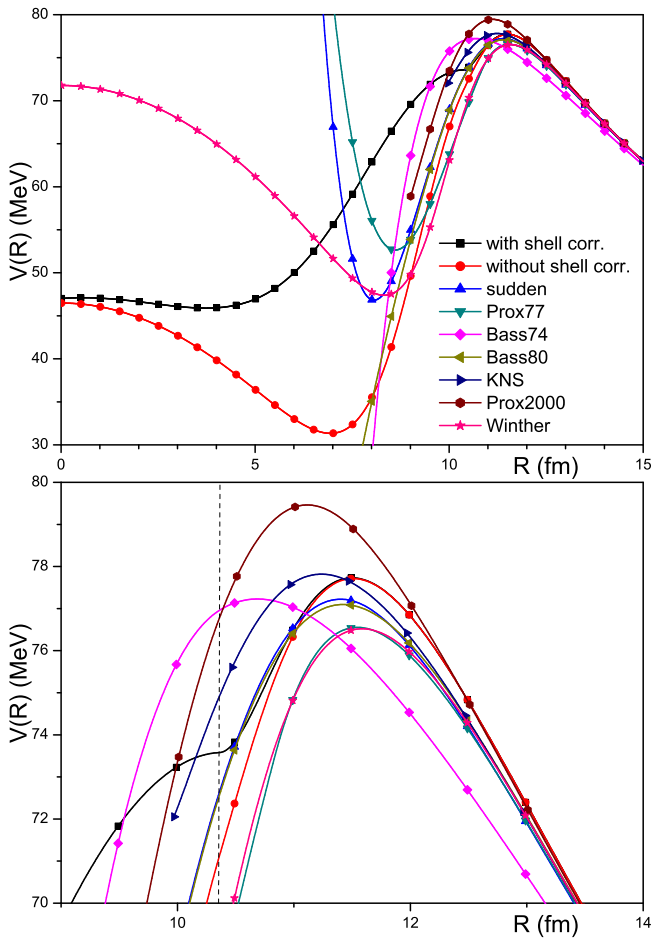}
\caption{(Color online) Full and macroscopic potentials for system $^{16}$O+$^{208}$Pb evaluated in my approach. The sudden-density (sudden) \cite{d2002}, Krappe-Nix-Sirk (KNS) \cite{kns}, Bass (Bass80) \cite{bass80}, Bass (Bass74) \cite{bass73}, proximity (Prox77) \cite{prox77}, proximity (Prox2000) \cite{prox2000} and Winther \cite{winther} potentials are presented too. Upper panel show the potentials in a wide range, while the bottom panel show the potentials around barriers. The vertical dashed line on the bottom panel shows the touching distance $R_t$.}
\end{center}
\end{figure}

The thicknesses of barrier at deep sub-barrier collision energies become larger due to the positive shell-correction contribution to the full potential, see Fig. 2. This directly leads to a reduction of barrier penetrability and, as the result, to the hindrance of deep sub-barrier fusion.

I compare these full and macroscopic potentials with other potentials, which are often used for analyzing of various heavy-ion reactions around the barrier, in Fig. 2. The shapes of Krappe-Nix-Sirk \cite{kns}, Bass80 \cite{bass80}, proximity \cite{prox77}, Winther \cite{winther}, sudden \cite{d2002} and macroscopic potentials are similar around barriers. In contrast to this, the full potential at below barrier collision energies, when the shell-correction contribution acts fully, is strongly deviated from the other ones. The shape of the full potential presented on the bottom panel of Fig. 2 is similar to the ones obtained in Refs. \cite{umar,hagino}.

\begin{figure}
\begin{center}
\includegraphics[width=8.7cm]{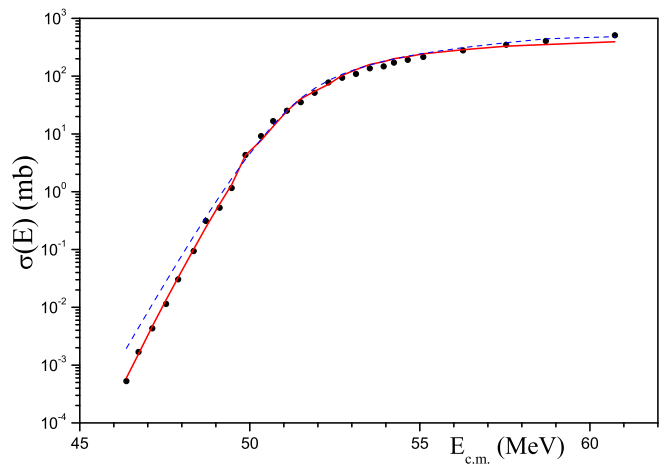}
\caption{(Color online) Fusion cross sections for reactions $^{48}$Ca+$^{48}$Ca. The experimental data are taken from Ref. \cite{CaCa} and the notation  is the same as in Fig. 1.}
\end{center}
\end{figure}

Comparison of my calculations of fusion cross sections for reactions $^{48}$Ca+$^{48}$Ca and $^{58}$Ni+$^{54}$Fe with experimental data \cite{CaCa,NiFe} are presented in Figs. 3 and 4, respectively. These data are well described in my approach, when the full nucleus-nucleus potential is taken into account. The values of shell-correction energy in $^{48}$Ca, $^{58}$Ni and $^{54}$Fe are $-2.648$, $-4.453$,  and $-3.840$ MeV, respectively. The values of potential parameters for these reactions are given in Table 1. The values of potential parameters evaluated for different reactions vary in the narrow ranges, see Table 1.  Here I use the same value of parameter $a$ as the one for reaction $^{16}$O+$^{208}$Pb, see Eq. (7).

\begin{figure}
\begin{center}
\includegraphics[width=8.7cm]{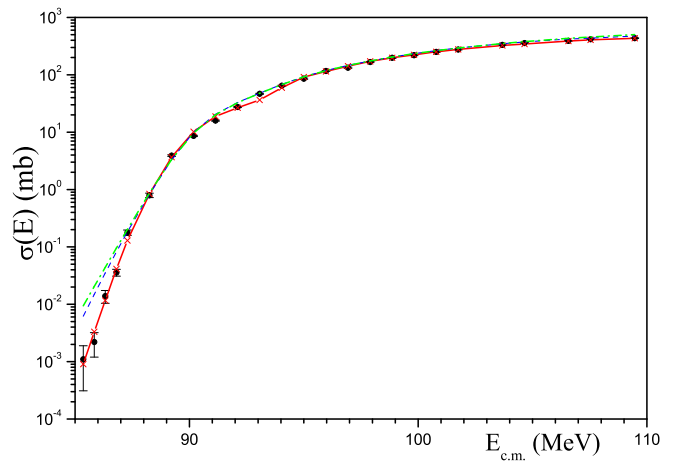}
\caption{(Color online) Fusion cross sections for the reaction $^{58}$Ni+$^{54}$Fe.
The experimental data are from Ref. \cite{NiFe}. The fusion cross sections obtained by using the macroscopic and shell-correction parts of the nucleus-nucleus potential are shown by the solid line, the fusion cross sections evaluated without shell-correction contribution to the potential is shown by the dashed line; and the fusion cross sections calculated with the opposite shell-correction contribution, by dash-dotted line.}
\end{center}
\end{figure}

Comparing the values of fusion cross sections near $1 \div 10 \; \mu$b obtained in various approaches in Figs. 1, 3--4 and the values shell-correction energies I conclude that colliding systems consisting of nuclei with larger absolute values of shell-correction energies have more prominent fusion hindrance at deep sub-barrier energies.

Note that some nuclei with numbers of proton and neutrons between the magic numbers or far from the $\beta$-stability line have positive values of the shell-correction energies. According to Eqs. (3) and (4) the shell-correction contribution to the potential at $\delta E_1 + \delta E_2 > 0$ reduces the thickness of the full potential compared to the thickness of the macroscopic potential. This enhances the barrier penetrability at low collision energies. Therefore the deep sub-barrier fusion cross section should be enhanced by the shell-correction contribution at $\delta E_1 + \delta E_2 > 0$ to the full potential.

To show this effect I changed the sign of the shell-correction energies in nuclei $^{58}$Ni and $^{54}$Fe to opposite sign and again evaluated of the fusion cross sections with the parameter values obtained earlier. The result of such calculation of the fusion cross sections for the system $^{58}$N+$^{54}$Fe are also presented in Fig. 4. Comparing the results of the different calculations in Fig. 4 we see weak enhancement of the deep sub-barrier fusion cross section induced by the positive shell-correction energies in colliding nuclei. The experimental investigation of this effect would be very interesting.

In conclusion, I have introduced the shell-correction contribution to the nucleus-nucleus potential and show that this contribution strongly influences the fusion cross section at deep sub-barrier energies. The macroscopic and full nucleus-nucleus potentials are different at distance $R \lesssim R_t$ due to the shell-correction contribution to the full potential.

\section*{Acknowledgments}

The author thanks Prof. M. Dasgupta for providing the cross section data for fusion reaction $^{16}$O+$^{208}$Pb, Prof. K. Hagino for useful discussion related to CCFULL code and Prof. B. Pritychenko for preprint of Ref. \cite{be2}.

\end{document}